\documentclass[aps,prd,superscriptaddress,showpacs,preprintnumbers]{revtex4}
\usepackage{graphicx,color,here,epsf,epstopdf,psfig,amsmath}
\usepackage[abs]{overpic}

\begin{document}

\newcommand{\dlt}{\bigtriangleup}
\newcommand{\beq}{\begin{equation}}
\newcommand{\eeq}[1]{\label{#1} \end{equation}}
\newcommand{\insertplot}[1]{\centerline{\psfig{figure={#1},width=14.5cm}}}

\parskip=0.3cm


\title{Resonance production in Pomeron-Pomeron collisions at the LHC}

\author{R. Fiore}
\affiliation{Department of Physics, University of Calabria, National
Institute of Nuclear Physics, I-87036 Arcavacata di Rende, Cosenza, ITALY}

\author{L. Jenkovszky}
\affiliation{Bogolyubov Institute for Theoretical Physics (BITP),
Ukrainian National Academy of Sciences \\14-b, Metrologicheskaya
str., Kiev, 03680, UKRAINE}

\author{R. Schicker}
\affiliation{Physikalisches Institut, Im Neuenheimer Feld 226, 
Heidelberg University, 69120 Heidelberg, GERMANY}

\begin{abstract}
A Regge pole model for Pomeron-Pomeron total cross section in the resonance
region \mbox{$\sqrt{M^2}\le$ 5 GeV} is presented. The cross section is 
saturated by direct-channel contributions from the Pomeron as well as from 
two different $f$ trajectories, accompanied by the isolated f$_0(500)$ 
resonance which dominates the $\sqrt{M^{2}}\lesssim 1$ GeV region. 
A slowly varying background is taken into account. The calculated 
Pomeron-Pomeron total cross section cannot be measured directly, 
but is an essential part of central diffractive processes. In preparation 
of future calculations of central resonance production at the hadron level, 
and corresponding measurements at the LHC, we normalize the Pomeron-Pomeron
cross section at large masses $\sigma_{t}^{PP} (\sqrt{M^2}\rightarrow \infty) 
\approx$ 1 mb as suggested by QCD-motivated estimates.
     
\end{abstract}


\maketitle

\vspace{-1.0cm}
\section{Introduction} \label{s1}
\vspace{-.6cm}
Central production in proton-proton collisions has been studied from the low 
energy range $\sqrt{s}$ = 12.7-63 GeV at the  ISR at CERN up to the presently 
highest energy of $\sqrt{s}$ = 13 TeV available in Run II at the LHC. Ongoing 
data analysis of central production events include data taken by the COMPASS 
Collaboration at the SPS \cite{COMPASS}, the CDF Collaboration at the 
TEVATRON \cite{CDF}, the STAR Collaboration at RHIC \cite{STAR},
and the ALICE and LHCb Collaborations at the LHC \cite{ALICE,LHCb}.  
A comprehensive survey of central exclusive production is given in a
recent review article \cite{Albrow1}.

The analysis of central production necessitates the simulation of such events 
to study the acceptance and efficiency of the complex large detector systems. 
With the existing detector upgrade programmes for central production 
measurements at RHIC and at the LHC, much larger data samples are expected in 
the next few years which will allow the analysis of differential distributions.
The purpose of the study  presented here is the development of a Regge pole 
model for simulating such differential distributions.

The study of central production, in particular at the soft scale, is 
interesting for a variety of reasons. Here, we refer to central production as 
arising from the fusion of two stronlgy interacting colour singlet objects, and
do not discuss any contributions due to photon exchange. The absence of a hard 
scale precludes a perturbative QCD description. The traditional framework for 
studying soft hadronic processes has been the Regge formalism. In this 
formalism, bound states are associated to Regge trajectories.
The classification of mesons by means of nonlinear Regge trajectories
has spectroscopic value by its own. At high energies, the hadronic interaction 
is dominated by the exchange of a leading trajectory, the pomeron. Within QCD, 
it is conjectured that this trajectory represents the exchange of purely 
gluonic objects. The study of central production at high energies allows to 
identify the contribution from the Pomeron trajectory. The dynamics of the 
corresponding multi-gluon colour-singlet exchange is presently only poorly 
understood within QCD, and such studies will hence contribute to an improved 
QCD-based understanding of Regge phenomenology. The fusion of multi-gluon 
objects is characterized by a gluon-dominated environment with highly 
suppressed quark degrees of freedom, and the evolution of this initial state 
is expected to populate with increased probability gluon-rich 
hadronic states, glueballs, and hybrids. The analysis 
of these centrally produced resonances by a Partial Wave Analysis reveals the 
quantum numbers $J^{PC}$ of these resonances. Of particular interest is the 
search for states with exotic quantum numbers which cannot be 
$q\bar{q}$-mesons, and hence must be exotic such as of tetra-quark nature 
($q\bar{q}$ + $\bar{q}q$), or gluonic hybrid ($q\bar{q} + gluon$). 
Moreover, the decomposition into states of known quantum numbers will shed  
new light also on the existence of numerous states in the scalar sector, 
a topic of fundamental interest \mbox{in hadron spectroscopy \cite{Ochs}.}

This article is organized as follows. In the introduction in Sect. 1,
the study of central production at hadron colliders is motivated. 
In Sect. 2, central production is reviewed. In Sect. 3, the dual resonance
model of Pomeron-Pomeron scattering is analysed. Non-linear complex meson 
trajectories are introduced in Sect. 4. Two $f$ trajectories, 
relevant for the calculation of the Pomeron-Pomeron 
cross section, are discussed in Sect. 5, while in Sect. 6 the 
Pomeron trajectory is presented. In Sect. 7, the f$_{0}$(500) resonance
is examined. The Pomeron-Pomeron total cross section is investigated
in Sect. 8. A summary and an outlook for more detailed studies
of the topic presented here \mbox{is given in Sect. 9.}
The procedure for fitting nonlinear complex meson trajectories 
is illustrated  in Appendix A for the example of the $\rho$-a trajectory.

\section{Central production}

Central production in proton-proton collisions is characterized 
by the two forward scattered protons, or remnants thereof, 
and by secondary particles produced at or close to mid-rapidity.
No particles are produced in the range between the mid-rapidity system 
and the two beam rapidities on either side of the central system.
Experimentally, these event topologies can be recognized by identifying 
the presence of the two rapidity gaps, by detecting the forward proton 
or its remnants, or by a combination of these two approaches.
Forward scattered neutral fragments can, for example, be detected 
in Zero Degree Calorimeters.

\begin{figure}[h]
\begin{center}
\begin{overpic}[width=.5\textwidth]{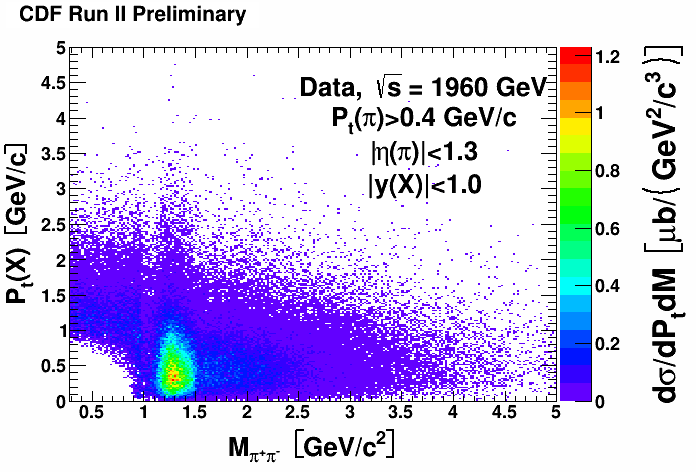}
\end{overpic}
\end{center}
\vspace{-.4cm}
\caption{Double differential cross section for central production of pion
pairs measured by the CDF Collaboration (Figure taken from Ref. \cite{Zurek}).}
\label{fig1}
\end{figure}

In Fig. \ref{fig1}, the differential cross section d$\sigma$\,/\,dP$_t$\,dM 
is shown for exclusive pion pair production in the CDF Run II 
at the TEVATRON. Clearly seen in this picture is the resonance
structure associated to the f$_2$(1270).
The complete kinematical determination of  the final state of centrally
produced pion or kaon pairs requires the measurement of the 3-momentum
(p$_x$,p$_y$,p$_z$) of both the positive and the negative partner of the pair.
The experimental single track acceptance is, however, limited by finite 
detector coverage in pseudorapidity, as well as by a cut-off in minimum 
transverse momentum p$_T$. This single track acceptance translates into 
missing acceptance for pairs of low mass and low transverse momentum. This 
missing pair acceptance is visible in Fig. \ref{fig1} in the data analyzed 
by the CDF Collaboration. For pairs of low masses, only the high end tail of 
the transverse momentum distribution can be measured. The 
extrapolation of the cross section to low transverse momenta is, however, 
possible based on models which are able to reproduce the resonance structures 
in the part of phase space covered by the detector acceptance.   
 
\begin{minipage}[t]{.29\textwidth}
\begin{overpic}[width=.98\textwidth]{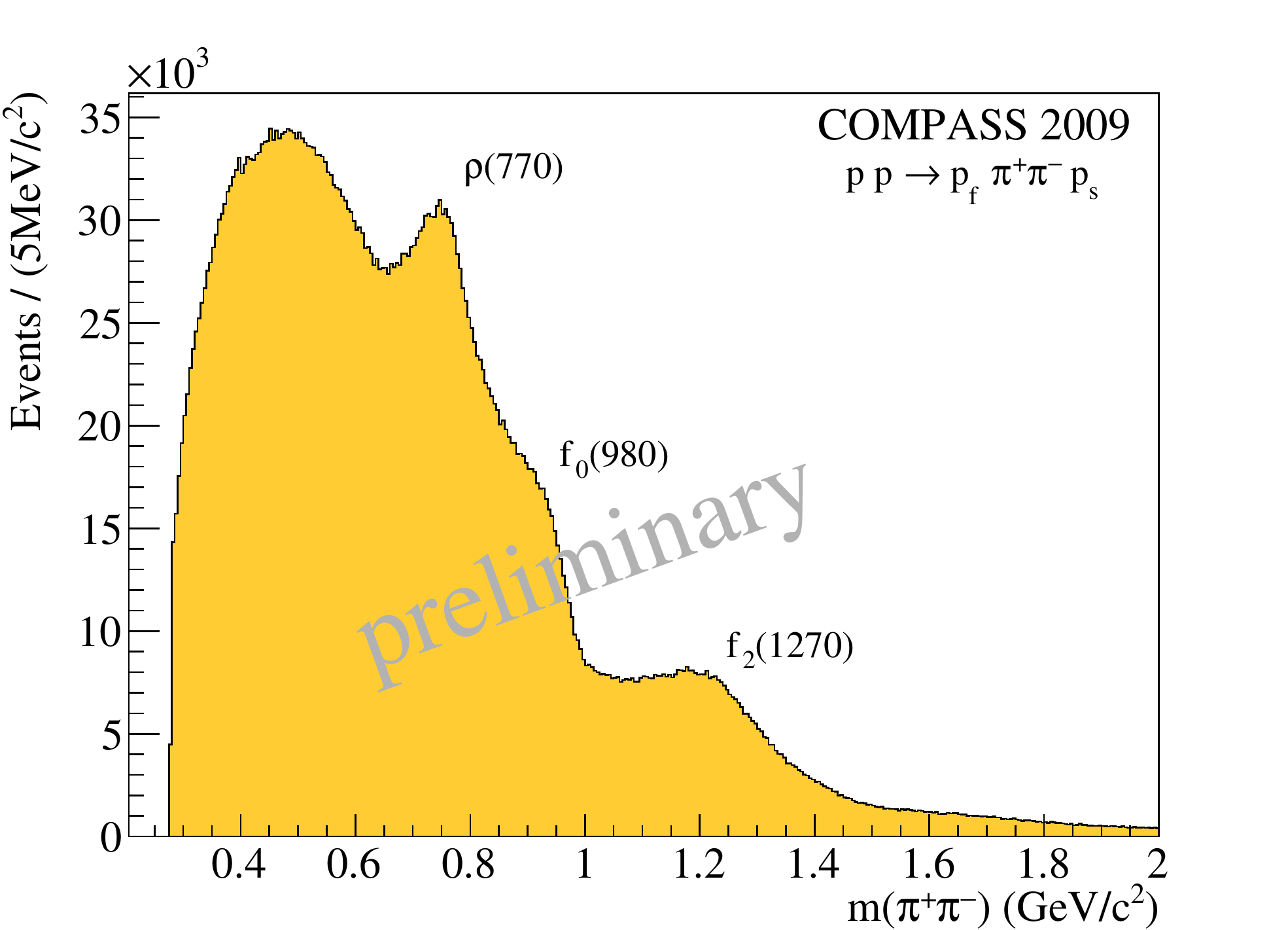}
\end{overpic}
\end{minipage}
\hspace{-0.2cm}
\begin{minipage}[h]{.36\textwidth}
\vspace{-2.9cm}
\begin{overpic}[width=.98\textwidth]{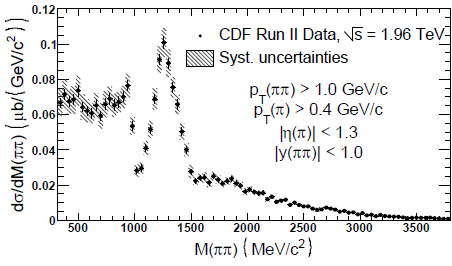}
\end{overpic}
\end{minipage}
\hspace{0.4cm}
\begin{minipage}[h]{.31\textwidth}
\vspace{-3.0cm}
\begin{overpic}[width=.98\textwidth]{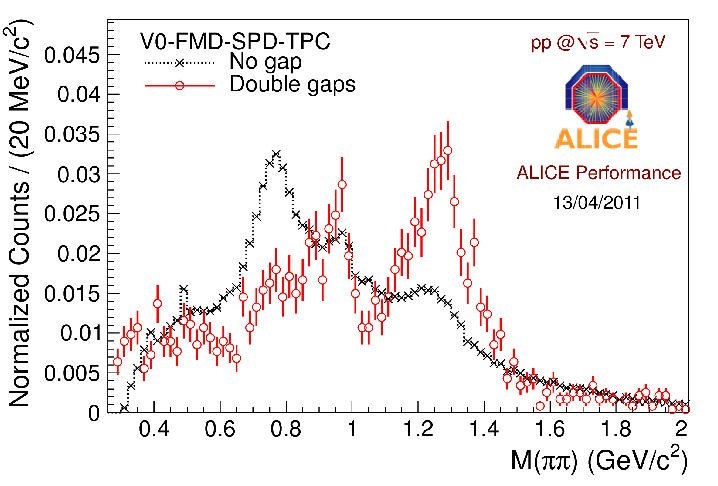}
\end{overpic}
\end{minipage}
\vspace{-.8cm}
\begin{figure}[h]
\includegraphics[width=.001\textwidth]{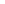}
\caption{Invariant pion pair masses from COMPASS Collaboration
on the left \cite{COMPASS_pair}, 
CDF Collaboration in the middle \cite{CDF_pair}, 
and ALICE Collaboration on the right \cite{ALICE_pair}.}
\label{fig2}
\end{figure}


The distributions of pion-pair invariant masses measured in proton-proton
collisions by the COMPASS, the CDF and the ALICE Collaboration are shown in 
Fig. \ref{fig2} on the left, in the middle and on the right, respectively.
In all these measurements, clear resonance structures are seen.
At the energy of the COMPASS measurement \mbox{$\sqrt{s}$ = 18.9 GeV,} 
Reggeons still contribute significantly to central production
as evidenced in the prominent $\rho$-peak. In addition, peaks associated 
to the f$_0$(980) and the f$_2$(1270) are seen, with a broad continuum 
extending to the two-pion threshold. At the higher TEVATRON energy of 
\mbox{$\sqrt{s}$ = 1.96 TeV,} the distribution at pair masses 
M $<$ 900 MeV/c$^2$ is significantly affected by the p$_{T}$ dependence 
of the acceptance as shown in Fig. \ref{fig1}. 
Full acceptance down to p$_T$ = 0  is reached 
\mbox{for masses M $>$ 900 MeV/c$^2$.} 
A clear resonance structure consistent with the f$_2$(1270) is seen at around
1270 MeV/c$^2$. A similar distribution is measured by the ALICE Collaboration 
at the LHC energy  $\sqrt{s}$ = 7 TeV as shown in Fig. \ref{fig2}
on the right. In addition, the ALICE double gap measurement is compared
to the pion-pair invariant mass distribution from no-gap events, i.e.
from inclusive production. In inclusive production, the $\rho$ as well as 
the K$_{s}^{0}$-signal are seen. These two signals are absent in the double-gap 
events, corroborating Pomeron-Pomeron dominance at the 
LHC energies discussed below.

\begin{figure}[h]
\begin{minipage}[t]{.98\textwidth}
\begin{center}
\begin{overpic}[width=.316\textwidth]{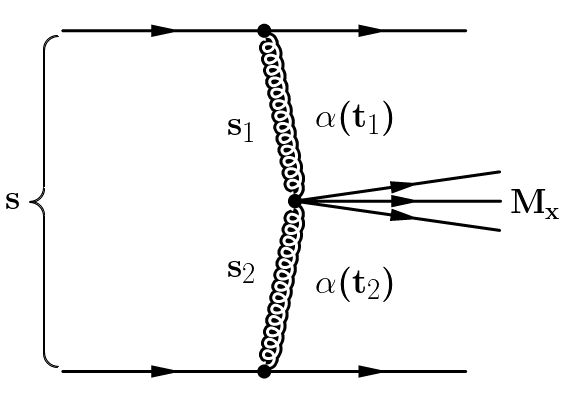}
\end{overpic}
\hspace{.7cm}
\begin{overpic}[width=.286\textwidth]{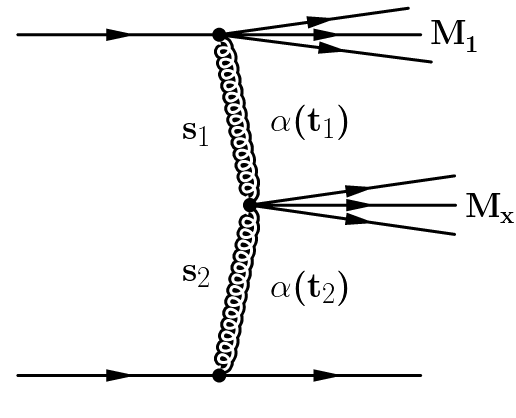}
\end{overpic}
\hspace{.7cm}
\begin{overpic}[width=.28\textwidth]{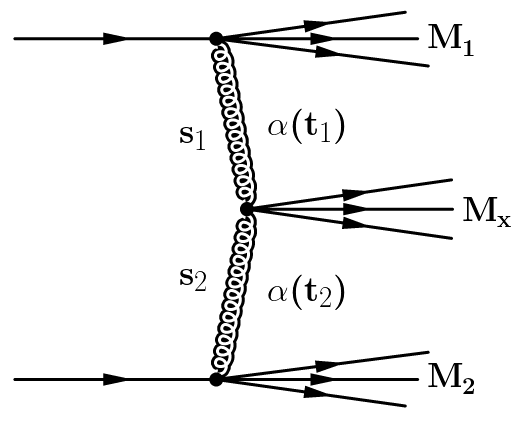}
\end{overpic}
\end{center}
\end{minipage}
\caption{Central production event topologies.}
\label{fig3}
\end{figure}

The double-gap topology of central production and the relevant kinematics are
shown in Fig. \ref{fig3}. This figure shows central production when the 
incoming protons remain in the ground state on the left, when one of the 
protons gets diffractively excited in the middle, and when both protons get 
excited on the right. All these reactions proceed by the exchange of Regge 
trajectories $\alpha(t_1)$ and $\alpha(t_2)$ which collide in the central 
region to produce a system of \mbox{mass M$_{\text x}$.} The total 
energy $s$ of the reaction is shared by the subenergies $s_1$ and $s_2$ 
associated to the trajectories $\alpha(t_1)$ and $\alpha(t_2)$, 
respectively. The LHC energies of $\sqrt{s}$ = 7, 8 and 13 TeV are 
sufficient to provide Pomeron dominance and allow for the neglect of 
Reggeon exchange which was not the case at the energies of previous 
accelerators. 

\begin{figure}[h]
\begin{center}
\begin{overpic}[width=.3\textwidth]{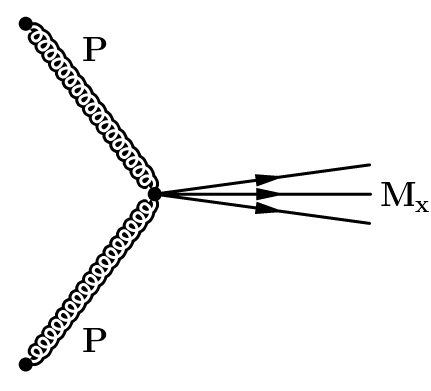}
\end{overpic}
\end{center}
\caption{Pomeron-Pomeron scattering.}
\label{fig4}
\end{figure}

The scope of the study presented here is the central part of the diagrams
shown in Fig. \ref{fig3}, i.e. Pomeron-Pomeron scattering producing 
mesonic states of mass M$_{\text x}$. We isolate the Pomeron-Pomeron-meson 
vertex shown in Fig. \ref{fig4}, and calculate the Pomeron-Pomeron total cross
section as a function of the centrally produced system of mass M$_{\text x}$.
The emphasis in this study is the behaviour in the low mass resonance region 
where perturbative QCD approaches are not applicable. 
Instead, similar to \cite{Jenk1,Jenk2}, we use the pole decomposition of a dual 
amplitude with relevant direct-channel trajectories $\alpha(M^2)$ for 
fixed values of Pomeron virtualities, $t_1=t_2=const.$ 
Due to Regge factorization, the calculated Pomeron-Pomeron cross section 
will enter the measurable proton-proton cross section \cite{Jenk3}.  

The nature of the Pomeron exchange is of fundamental interest for QCD-based
studies of exchange amplitudes. An effective vectorial-exchange is very
successful in reproducing the energy dependence of hadron-hadron cross
sections \cite{DL}. Such an approach results in opposite signs for proton-proton
and proton-antiproton amplitudes. Pomeron exchange, however, must yield
the same sign for these two reaction channels. Recent studies
on soft high-energy scattering solve this problem in terms of
effective propagators and vertices for the Pomeron exchange \cite{Ewerz}.
Within this model, the Pomeron exchange is decribed as an effective
rank-two tensor exchange \cite{Lebie,Bolz}. 

\setlength{\textheight}{250mm}
\section{Dual resonance model of Pomeron-Pomeron scattering}

The study of Pomeron-Pomeron $(PP)$ scattering is related to photon-photon 
scattering, the main difference being the positive and negative C-parity 
of the Pomeron and photon, respectively. 
High-virtuality $\gamma^*\gamma^*$ scattering is a favourite process in the 
framework of perturbative QCD, where the total cross 
section was calculated in Ref. \cite{Bartels}. 
In the leading-order BFKL
\begin{equation} \label{eq:QCD}  
\sigma_{tot}^{\gamma^*\gamma^*}=\frac{\sigma_0}{\sqrt{Q_1^2Q_2^2Y}}(s/s_0)^\lambda,
\end{equation}
where $Q^2_i=-q^2_i,\ \ i=1,2$ is the photon virtuality and 
$Y=\ln\bigl(\frac{s}{s_0}\bigr)$. The quantity $\sigma_0$ is a free parameter 
and the exponent $\lambda\equiv\alpha_P^{BFKL}$ is the familiar BFKL eigenvalue 
$N_c\alpha_s4\ln 2/\pi.$  We recall that the transition  
from photon-photon to  central Pomeron-Pomeron scattering is accompanied by
the change of variables $Q^2\rightarrow t$ and $s\rightarrow M^2$.
The above result was improved in Refs. \cite{Brodsky,Caporale}.

Most of the studies on diffraction dissociation, single, double and central, 
use the triple Reggeon formalism. This approach is useful in the smooth Regge 
region, beyond the resonance region, but is not applicable for the 
production of low masses which is dominated by resonances. We solve this 
problem by using a dual model. 

The  one-by-one account of single resonances is possible, but not 
economic for the calculation of cross section, to which a sequence of many 
resonances contributes at low masses. These resonances overlap and gradually 
disappear in the continuum at higher masses. An approach to account for many 
resonances, based on the idea of duality with a limited number of resonances 
lying on nonlinear Regge trajectories, was suggested in Ref. \cite{JLAB}. 
Later on, this approach was used in Refs. \cite{Jenk1,Jenk2} to calculate low 
mass single- and double-diffractive dissociation at the LHC.  

\begin{figure}[htb]
\includegraphics[width=.19\textwidth]{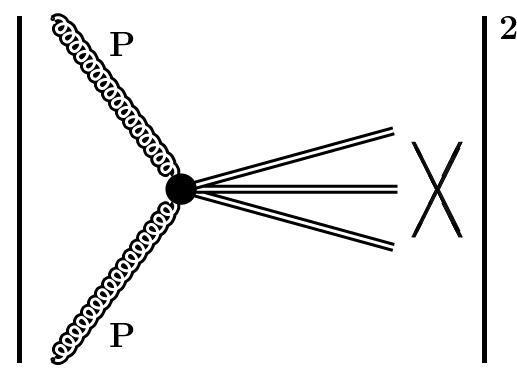}
\includegraphics[width=.038\textwidth]{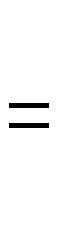}
\hspace{-0.2cm}
\includegraphics[width=.154\textwidth]{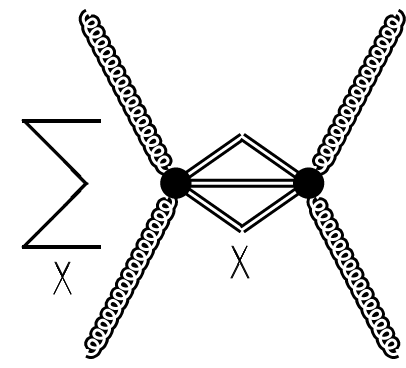}
\hspace{0.2cm}
\includegraphics[width=.037\textwidth]{pp0_0.png}
\begin{overpic}[width=.12\textwidth]{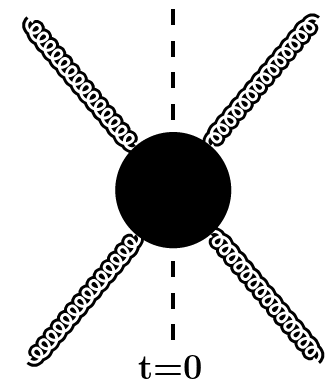}
\put(-34.,20.){\bf Unitarity}
\end{overpic}
\hspace{-0.2cm}
\includegraphics[width=.037\textwidth]{pp0_0.png}
\includegraphics[width=.132\textwidth]{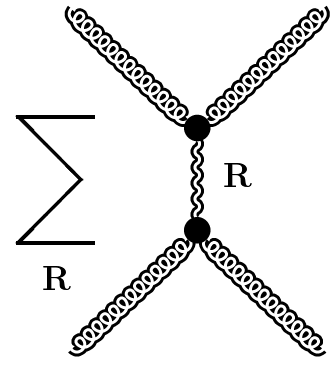}
\hspace{0.2cm}
\includegraphics[width=.038\textwidth]{pp0_0.png}
\hspace{0.2cm}
\begin{overpic}[width=.14\textwidth]{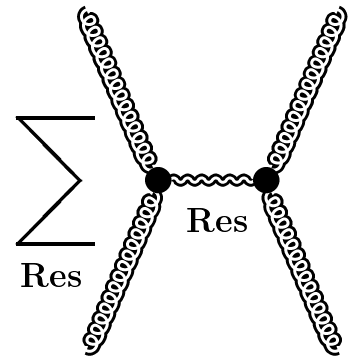}
\put(-50.,20.){\bf Veneziano }
\put(-38.,10.){\bf duality }
\end{overpic}
\caption{Connection, through unitarity (generalized optical
theorem) and Veneziano-duality, between the Pomeron-Pomeron cross section 
and the sum of direct-channel resonances.}
\label{fig5}
\end{figure}

The main idea behind this approach is illustrated in Fig. \ref{fig5}, 
realized by dual amplitudes with Mandelstam analyticity (DAMA) \cite{DAMA}.
For $s\rightarrow\infty$ and fixed $t$ it is Regge-behaved.
Contrary to the Veneziano model, DAMA not only allows
for, but rather requires the use of nonlinear complex trajectories
providing the resonance widths via the imaginary part of the
trajectory. In the case of limited real part of the
trajectory, a finite number of resonances is produced. More
specifically, the asymptotic rise of the trajectories in DAMA is
limited by the condition, in accordance with an important upper
bound, 

\vspace{-0.6cm}
\begin{eqnarray}
 |{\alpha(s)\over{\sqrt s\ln s}}|\leq const, \ \
s\rightarrow\infty.
\label{eq1}
\end{eqnarray}

For our purpose of central production, the direct-channel pole decomposition 
of the dual amplitude $A(M_{X}^{2},t)$ is relevant. Different trajectories 
$\alpha_{i}(M_X^2)$ contribute to this amplitude, with $\alpha_{i}(M_X^2)$ a 
nonlinear, complex Regge trajectory in the Pomeron-Pomeron system,

\vspace{-0.6cm}
\begin{eqnarray}
A(M_X^2,t)=a\sum_{i=f,P}\sum_{J}\frac{[f_{i}(t)]^{J+2}}{J-\alpha_i(M_X^2)}.
\label{eq2}
\end{eqnarray}
  
The pole decomposition of the dual amplitude  $A(M_{X}^{2},t)$ is shown 
in Eq. (\ref{eq2}), with $t$ the squared momentum transfer in the 
$PP\rightarrow PP$ reaction. The index $i$ sums over the trajectories which 
contribute to the amplitude. Within each trajectory, the second sum extends 
over the bound states of spin $J$. The prefactor $a$ in Eq. (\ref{eq2}) has 
the numerical value a = 1 GeV$^{-2}$ = 0.389 mb.

The pole residue $f(t)$ appearing in the $PP\rightarrow PP$ system
is fixed by the dual model, in particular by the compatibility of its Regge
asymptotics with Bjorken scaling and reads
\begin{equation}
\label{eq3} f(t)=(1-t/t_0)^{-2},
\end{equation}

where $t_0$ is a parameter to be fitted to the data. However, due to
the absence of data so far, we set $t_0=0.71$ GeV$^2$ for the moment
as in the proton elastic form factor. Note that the residue enters with a
power ($J\!+\!2$) in \mbox{Eq. (\ref{eq2}),} thereby strongly damping higher spin 
resonance contributions.

The imaginary part of the amplitude $A(M_X^2,t)$ given in Eq. (\ref{eq2})
is defined by 

\begin{equation} \label{ImA}
\Im m\, A(M_{X}^2,t)=a\sum_{i=f,P}\sum_{J}\frac{[f_{i}(t)]^{J+2} 
\Im m\,\alpha_{i}(M_{X}^2)}{(J-Re\,\alpha_{i}(M_{X}^2))^2+ 
(\Im m\,\alpha_{i}(M_{X}^2))^2}.
\end{equation}

For the $PP$ total cross section we use the norm 

\vspace{-0.3cm}
\begin{eqnarray}
\sigma_{t}^{PP} (M_{X}^2)= {\Im m\; A}(M_{X}^2, t=0),
\label{eq:ppcross}
\end{eqnarray}

and recall that the amplitude $A$ and the cross section $\sigma_{t}$ carry 
dimensions of mb due to  \mbox{the dimensional parameter $a$} discussed above.
The Pomeron-Pomeron channel, $PP\rightarrow M_X^2$, couples to the Pomeron
and $f$ channels dictated by conservation of quantum numbers.
In order to calculate the $PP$ cross section, we therefore take into account
the trajectories associated to the f$_0$(980) and to the f$_2$(1270) resonance, 
and the Pomeron trajectory.

\section{Nonlinear, complex meson Regge trajectories}
\label{Sec:trajectory}

A non-trivial task for analytic models of Regge trajectories consists
in deriving the imaginary part of the trajectory from the seemingly linearly
increasing real part \cite{Degasperis,Fiore1,Fiore2}.
The importance of the nonlinearity of the real part was studied
in Refs. \cite{Filipponi,Fiore,Brisudova}. A dispersion relation connects 
the real and imaginary part of the trajectory.

We follow Ref. \cite{Fiore1} to relate the nearly linear real part of 
the meson trajectory to its imaginary part,  

\begin{eqnarray}
\Re e\:\alpha(s) = \alpha(0) + \frac{s}{\pi} 
PV \int_0^{\infty} ds^{'} \frac{\Im m\:\alpha(s^{'})}{s^{'}(s^{'}-s)}. 
\label{eq:disp}
\end{eqnarray}

In Eq. (\ref{eq:disp}), PV denotes the Cauchy Principal Value of the integral.
The imaginary part  is related to the decay width by 

\begin{eqnarray}
\Gamma(M_{R}) = \frac{\Im m\: \alpha(M_{R}^{2})}{\alpha^{'}\:M_{R}}.
\label{eq:width}
\end{eqnarray}

The quantity $\alpha^{'}$ in Eq. (\ref{eq:width}) denotes the derivative of the 
real part, $\alpha^{'}$ = $\frac{d \Re e\:\alpha(s)}{ds}$. The relation 
between $\Gamma(M)$ and $\Im m\:\alpha$(s) requires $\Im m\:\alpha(s) > 0$. 
In a simple analytical model, the imaginary part is chosen as a 
sum of single threshold terms \cite{Fiore1}

\begin{eqnarray}
\Im m\: \alpha(s) = \sum_{n} c_{n} (s-s_{n})^{1/2} 
\big(\frac{s-s_{n}}{s}\big)^{|\Re e\:\alpha(s_{n})|} \theta(s-s_{n}).
\label{eq:imag}
\end{eqnarray}

The imaginary part of the trajectory shown in Eq. (\ref{eq:imag})
has the correct threshold and asymptotic behaviour. 
Since $\Im m\:\alpha(s) > 0 $, all the expansion coefficients $c_{\text n}$
must be positive. The values of s$_{\text n}$ represent kinematical thresholds of 
decay channels. The highest threshold, higher than all the resonance
masses lying on the trajectory, is chosen as an effective threshold.
This highest threshold ensures that $\Re e\:\alpha$(s) tends to a constant
value for  $s \rightarrow \infty$. 

The parameterisation of the real and imaginary part of a meson trajectory, and
the extraction of the expansion coefficients c$_n$ shown in Eq. (\ref{eq:imag}),
are derived in Appendix A for the case of the $\rho$-a trajectory.

\setlength{\textheight}{270mm}

\section{Two $f$ trajectories}

\vspace{-.2cm}
Apart from the Pomeron trajectory discussed below, the direct-channel 
$f$ trajectory is essential in the PP system. Guided by conservation
of quantum numbers, we include two $f$ trajectories, labelled 
$f_1$ and $f_2$, with mesons lying on these trajectories 
as specified in Table \ref{table1}.

\begin{center}
\begin{table}[h]
\begin{tabular}{| c | c c | c || c | c | c ||}
\hline
& I$^{G} $& J$^{PC}$ & traj. & M (GeV) & M$^{2}$ (GeV$^{2}$) &  $\Gamma$ (GeV) \\ 
\cline{1-7}
f$_{0}$(980) & 0$^{+}$ &0$^{++}$ & $f_{1}$ &0.990$\pm$0.020 &0.980$\pm$0.040 &0.070$\pm$ 0.030\\ 
f$_{1}$(1420) & 0$^{+}$ &1$^{++}$ & $f_{1}$ &1.426$\pm$0.001 &2.035$\pm$0.003 &0.055$\pm$ 0.003\\ 
f$_{2}$(1810) & 0$^{+}$ &2$^{++}$ & $f_{1}$ &1.815$\pm$0.012 &3.294$\pm$0.044 &0.197$\pm$ 0.022\\ 
f$_{4}$(2300) & 0$^{+}$ &4$^{++}$ & $f_{1}$ &2.320$\pm$0.060 &5.382$\pm$0.278 &0.250$\pm$ 0.080\\ 
f$_{2}$(1270) & 0$^{+}$ &2$^{++}$ & $f_{2}$ &1.275$\pm$0.001 &1.6256$\pm$0.003 &0.185$\pm$ 0.003\\ 
f$_{4}$(2050) & 0$^{+}$ &4$^{++}$ & $f_{2}$ &2.018$\pm$0.011 &4.0723$\pm$0.044 &0.237$\pm$ 0.018\\ 
f$_{6}$(2510) & 0$^{+}$ &6$^{++}$ & $f_{2}$ &2.469$\pm$0.029 &6.096$\pm$0.143 &0.283$\pm$ 0.040\\ 
\hline
\end{tabular}   
\caption{Parameters of resonances belonging to the $f_1$ and $f_2$ trajectories.} 
\label{table1}
\end{table}
\end{center}

\vspace{-.6cm}
The real and imaginary part of the $f_{1}$ and $f_{2}$ trajectories can 
be derived as discussed in Appendix A from the parameters of the f-resonances 
listed in Table \ref{table1}.

\vspace{-.2cm}
\begin{figure}[h]
\begin{minipage}[t]{.9\textwidth}
\begin{center}
\begin{overpic}[width=.44\textwidth]{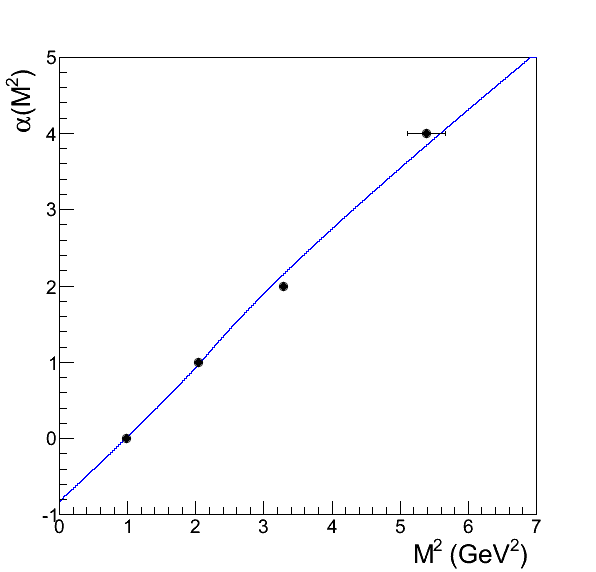}
\end{overpic}
\hspace{.2cm}
\begin{overpic}[width=.44\textwidth]{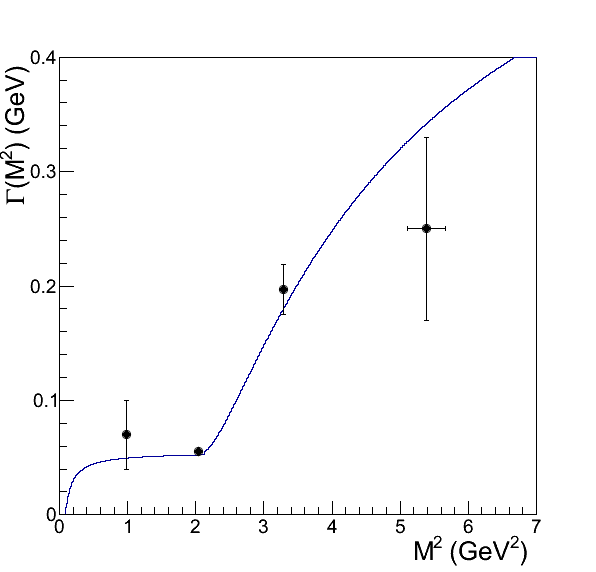}
\end{overpic}
\end{center}
\end{minipage}
\vspace{-0.1cm}
\caption{Real part of $f_{1}$ trajectory on the left, width function 
$\Gamma$(M$^{2}$) on the right.}
\label{fig:fig3}
\end{figure}

The real part and the width function of the $f_{1}$ trajectory are shown in 
Fig. \ref{fig:fig3} on the left and right, respectively. In the fit of 
this trajectory, the same three thresholds are used as discussed in Appendix A
for the $\rho$-a trajectory.

\vspace{-.3cm}
\begin{figure}[h]
\begin{minipage}[t]{.9\textwidth}
\begin{center}
\begin{overpic}[width=.44\textwidth]{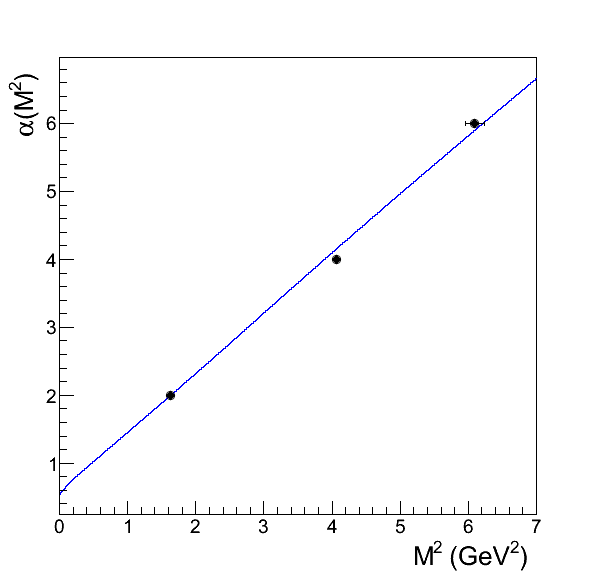}
\end{overpic}
\hspace{.2cm}
\begin{overpic}[width=.44\textwidth]{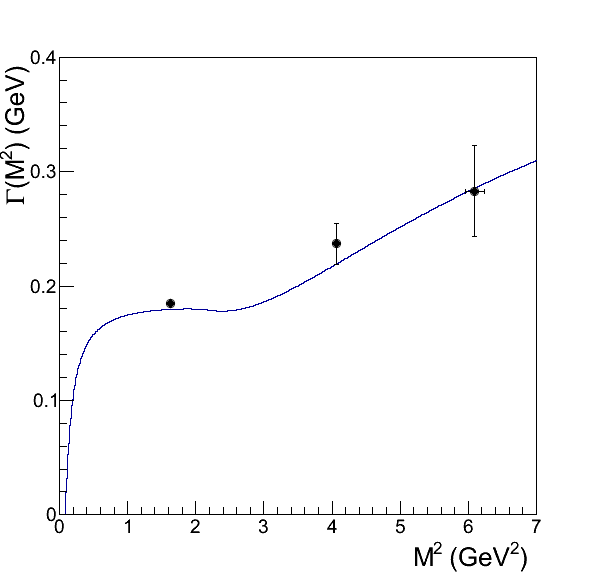}
\end{overpic}
\end{center}
\end{minipage}
\vspace{-0.1cm}
\caption{Real part of $f_{2}$ trajectory on the left, width function 
$\Gamma$(M$^{2}$) on the right.}
\label{fig:fig4}
\end{figure}

The real part and the width function of the $f_{2}$ trajectory are shown in 
Fig. \ref{fig:fig4} on the left and right, respectively. In the fit of 
this trajectory $f_{2}$, the same three thresholds are used as for the $f_{1}$
trajectory.

\setlength{\textheight}{240mm}

\section{The Pomeron trajectory}

While ordinary meson trajectories can be fitted both in the resonance and 
scattering region corresponding to positive and negative values of the 
argument, the parameters of the Pomeron trajectory can only be determined in 
the scattering region $M^2<0$. The poles of this trajectory at $M^2>0$ are 
identified with glueball candidates. An extensive literature on such candidates
exists, including theoretical predictions and experimental identification. 
The status of glueballs is, however, controversial and a topic of ongoing 
discussions and debate; see Refs. \cite{Ochs,Kirk} and references therein.
In this study, we associate the bound states of the Pomeron trajectory to
glueball candidates, as has been done previously
in Refs. \cite{DJL,Kaidalov,Brisudova,Sergienko}.  
  
A comprehensive fit to high-energy $pp$ and $p\bar{p}$ of 
the nonlinear Pomeron trajectory  is discussed in Ref.\cite{Jenk3}

\begin{eqnarray}
\alpha_P(M^2) = 1. + \varepsilon + \alpha^{'} M^2 - c\sqrt{s_{0}-M^2},
\label{eq:pom1}
\end{eqnarray}

with $\varepsilon$\,=\,0.08, $\alpha^{'}$\,=\,0.25 GeV$^{-2}$ and s$_0$ 
the two pion threshold s$_0$\,=\,4m$_{\pi}^{2}$. The value of c is taken as 
c\,=\,$\alpha^{'}$/10\,=\,0.025.

In order to be consistent with the mesonic trajectories shown above, the
linear term in Eq. (\ref{eq:pom1}) is replaced by a heavy threshold mimicking
linear behaviour in the mass region of interest (M $<$ 5 GeV),

\begin{eqnarray}
\alpha_{P}(M^2)=\alpha_0+\alpha_1\Big(2m_{\pi}-\sqrt{4m_{\pi}^2-M^2}\Big)
+\alpha_2\Big(\sqrt{M^2_H}-\sqrt{M^2_H-M^2}\Big),
\label{eq:pom2}
\end{eqnarray}

where $M_H$ is an effective heavy threshold set at  $M=3.5$ GeV. The
coefficients $\alpha_{0},\alpha_{1}$ and $\alpha_{2}$ 
are chosen such that the Pomeron trajectory of Eq. (\ref{eq:pom2}) 
has a low energy behaviour as defined by Eq. (\ref{eq:pom1}).

\begin{figure}[h]
\begin{minipage}[t]{.98\textwidth}
\begin{center}
\begin{overpic}[width=.48\textwidth]{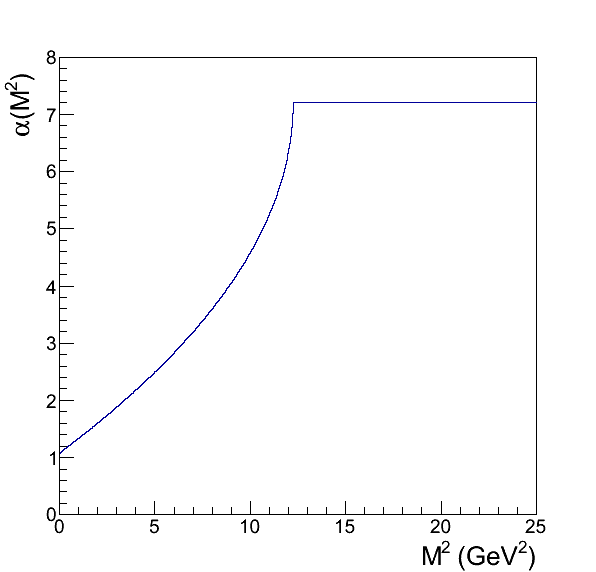}
\end{overpic}
\hspace{.5cm}
\begin{overpic}[width=.48\textwidth]{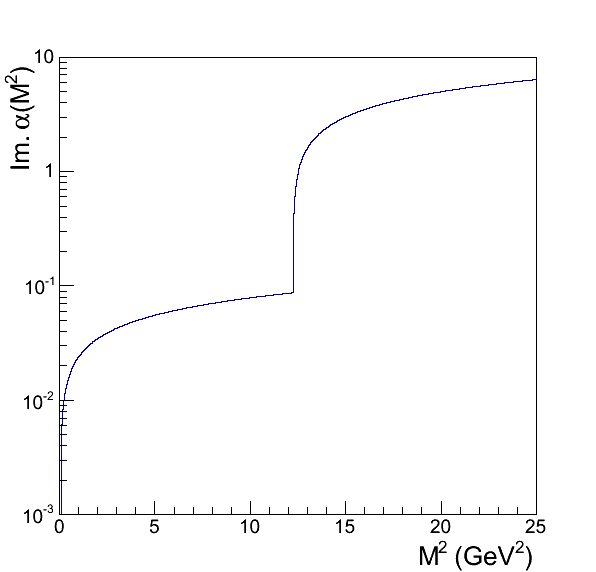}
\end{overpic}
\end{center}
\end{minipage}
\caption{Real part of pomeron trajectory on the left, 
imaginary part on the right.}
\label{fig:fig5}
\end{figure}

The real and imaginary part of the Pomeron trajectory resulting from the 
parameterisation of Eq. (\ref{eq:pom2})  is shown in Fig. (\ref{fig:fig5}) 
on the left and right, respectively. Clearly visible is the asymptotically 
constant value of the real part beyond the heavy threshold, accompanied by 
a strong increase of the imaginary part.

\section{The $f_{0}(500)$ resonance}

The experimental data on central exclusive pion-pair production measured at the 
energies of the ISR, RHIC, TEVATRON and the LHC collider all show a broad 
continuum for pair masses m$_{\pi^+\pi^-} <$ 1 GeV/c$^{2}$. This mass region is 
experimentally difficult to access due to the missing acceptance for 
pion-pairs of low mass and low transverse momentum p$_{T}$ as discussed above.
The population of this  mass region is attributed to the $f_{0}$(500), a 
resonance which has been controversial for many decades. In the 2010 edition 
of the Review of Particle Physics (RPP), this resonance is listed as 
$f_{0}$(600) with a mass M$_{0}$ in the range $400 < M_{0} < 1200\,$MeV,
and a width $\Gamma$ in the range $600 < \Gamma < 1000\,$MeV.
Since the RPP edition of 2012, this resonance is listed 
as $f_{0}$(500) with mass in the range $400 < M_{0} < 550\,$MeV,
and width in the range $400 < \Gamma < 700\,$MeV \cite{Pelaez}.

The $f_{0}$(500) resonance is of prime importance for the understanding 
of the attractive part of the nucleon-nucleon interaction, as well as for
the mechanism of spontaneous breaking of chiral symmetry. 
The nature of the $f_{0}$(500) is a topic of ongoing studies and discussions, 
it is however generally agreed that it cannot be interpreted as a predominant 
$q\bar{q}$-state. The non-ordinary nature of the $f_{0}$(500) resonance is  
corroborated by the fact that it does not fit into the Regge description of 
classifying $q\bar{q}$-states into trajectories \cite{Anisovich}. 
A possible interpretation of the $f_{0}$(500) is a tetra-quark
configuration consisting of two valence and two antiquarks
in  colour-neutral state. It was shown  that such a configuration can give 
rise to a nonet of light scalar-isoscalar mesons \cite{Jaffe}.
Different approaches interpret the $f_{0}$(500) as arising from 
an inner tetra-quark structure and changing to an outer structure 
of a pion-pion state \cite{Close}.  There is strong 
evidence that this $f_{0}$(500) state belongs to a SU(3) nonet composed of 
the $f_{0}$(500), $f_{0}$(980), $a_{0}$(980) and $K_{0}^{*}$(800).

In spite of the complexity of the $f_{0}$(500) resonance, and the controversy 
on its interpretation and description, we take here the practical but 
simple-minded approach of a Breit-Wigner resonance \cite{PDG} 

\begin{eqnarray} 
A(M^{2}) = a\; \frac{-M_0\Gamma}{M^{2}-M_{0}^{2}+iM_{0}\Gamma}. 
\label{eq:BWampl}
\end{eqnarray} 

In Eq. (\ref{eq:BWampl}), the parameterisation of the relativistic Breit-Wigner 
amplitude is shown with M$_{0}$ and $\Gamma$ the mass and width, respectively. 
Here, the prefactor $a$ is added for consistency with the definition
of the amplitude \mbox{shown in Eq. (\ref{eq2}).}  
The Breit-Wigner amplitude of Eq. (\ref{eq:BWampl}) is used below  for 
calculating the contribution of the $f_{0}$(500) resonance to the 
Pomeron-Pomeron cross section.

\section{Pomeron-Pomeron total cross section}

The Pomeron-Pomeron cross section is calculated from the 
imaginary part of the amplitude by use of the optical theorem

\vspace{-0.4cm}
\begin{eqnarray}
\sigma_{t}^{PP} (M^2) \;\; = \;\; {\Im m\; A}(M^2, t=0) \;\; =  \;\; 
\sum_{i=f,P}\sum_{J}\frac{[f_{i}(0)]^{J+2}\; \Im m \;\alpha_{i}(M^{2})}
{(J-\Re e \;\alpha_{i}(M^{2}))^{2}+(\Im m \;\alpha_{i}(M^{2}))^{2}}.
\label{eq:imampl}
\end{eqnarray}

\vspace{-0.2cm}
In Eq. (\ref{eq:imampl}), the index $i$ sums over the trajectories which
contribute to the cross section, in our case the $f_{1}$, $f_{2}$
and the Pomeron trajectory discussed above.
Within each trajectory, the summation extends over the bound states
of spin $J$ as expressed by the second summation sign.
The value $f_{i}(0) =f_{i}(t)\big |_{\text t=0}$ is not known a priori.
The analysis of relative strengths of the states of trajectory $i$ will, 
however, allow to extract a numerical value for $f_{i}$(0) from the 
experimental data .

The Breit-Wigner parameterisation of the isolated $f_{0}$(500) 
resonance contributes to the cross section with

\vspace{-0.4cm}
\begin{eqnarray} 
\sigma_{f_{0}(500)}^{PP}(M^{2}) = a \sqrt{1.-\frac{4\,m_{\pi}^{2}}{M^{2}}}
\frac{{M}_{0}^{2}\Gamma^{2}}{({M}^{2}-{M}_{0}^{2})^{2}+{M}_{0}^{2}\Gamma^{2}}, 
\label{eq:BWcross}
\end{eqnarray} 

with the resonance mass of M$_{0}$ = (0.40--0.55) GeV and a width 
$\Gamma$ = (0.40--0.70) GeV \cite{PDG}.  The quantity
$\sqrt{1.-4\,m_{\pi}^{2}/M^{2}}$ in Eq. (\ref{eq:BWcross}) is the 
threshold phase space factor for the two-pion decay.

In addition to the contributions discussed above, a background 
term is added  to the PP cross section. This background is of form
\cite{Kononenko}

\begin{eqnarray}
\sigma_{backgr.}^{PP}(M^{2}) = c*(0.1+ \text{log}(M^{2}))\text{\;mb},
\label{ew:back}
\end{eqnarray}

with the numerical value of the parameter c fitted to data.

\begin{figure}[h]
\begin{minipage}[t]{.99\textwidth}
\begin{overpic}[width=.99\textwidth]{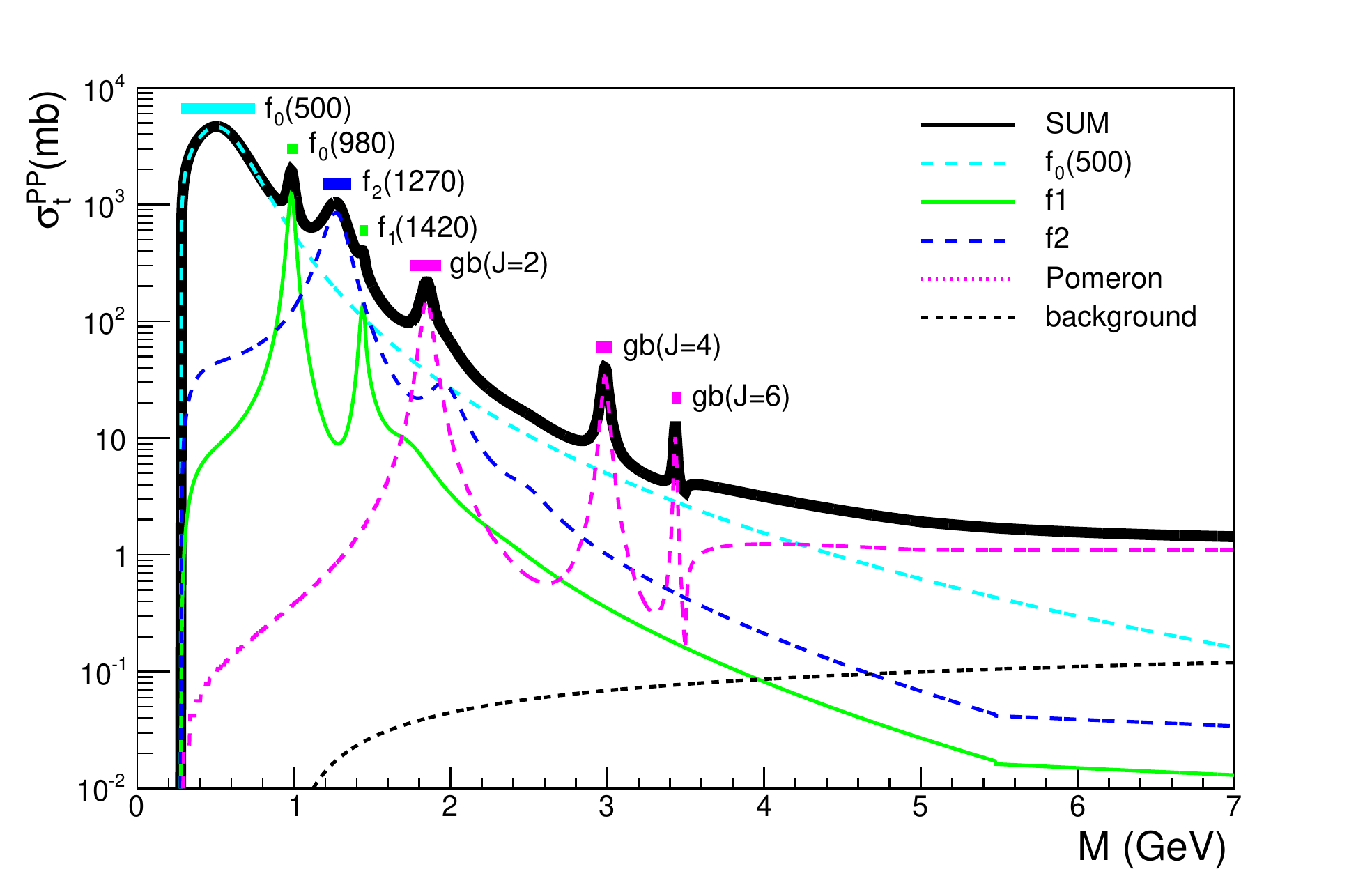}
\end{overpic}
\end{minipage}
\caption{Contributions of the f$_{0}$(500) resonance, the $f_{1}$, $f_{2}$ 
and the Pomeron trajectory, and of the background to PP total cross section.}
\label{fig:fig6}
\end{figure}

In Fig. \ref{fig:fig6}, the different contributions to the
PP total cross section are shown. The contribution of the 
$f_{0}$(500) resonance according to Eq. (\ref{eq:BWcross}) is displayed
by the dashed cyan line.
The contribution of the $f_{1}$ trajectory indicated by the 
solid green line  clearly shows  the f$_0$(980) and the 
f$_1$(1420) resonance. The higher mass states, the f$_2$(1810)
and the f$_4$(2300), are barely visible due to their reduced cross section 
and much larger width. Similarly, the contribution of the $f_{2}$
trajectory indicated by the dashed blue line shows peaks for  
the f$_2$(1270) and the f$_4$(2050) resonances, with the 
f$_6$(2510) barely visible. The contributions of both $f_1$ and $f_2$
trajectory show a kink at about M = 5.5 GeV due to the heavy
threshold s$_{2}$ = 30 GeV$^{2}$. The contribution from the Pomeron 
trajectory is displayed  in Fig. \ref{fig:fig6} 
by the dashed magenta line. Visible here is the resonance structure
due to the $J=2,\ 4$ and $6$ states on the trajectory labelled
by gb(J=2), gb(J=4) and gb(J=6), respectively. 
Beyond the heavy threshold, M = 3.5 GeV, the transition to the continuum 
is seen, reflecting the behaviour of the real and imaginary part of the 
trajectory as shown in Fig. \ref{fig:fig5}.
The background contribution to the PP cross section is shown in 
Fig. \ref{fig:fig6} by the dashed black line, and is normalized
here to represent approximately 10\% of the signal at M = 7 GeV.  

The  Pomeron-Pomeron total cross section is calculated by summing over the 
contributions discussed above, and is shown in Fig. \ref{fig:fig6}
by the solid black line. The prominent structures seen in the total cross
section are labelled by the resonances generating the peaks.
The model presented here does not specify the 
relative strength of the different contributions shown in 
Fig. \ref{fig:fig6}. A Partial Wave Analysis of experimental data on 
central production events will be able to extract the quantum numbers 
of these resonances, and will hence allow to associate each resonance
to its trajectory. The relative strengths of the  contributing
trajectories need to be determined from the experimental data.

\section{Summary and Outlook}

A Regge pole model is presented  for calculating the Pomeron-Pomeron total 
cross section in the resonance region $\sqrt{M^{2}} \le$ 5 GeV. The 
direct-channel contributions of the Pomeron and two $f$ trajectories, including
a background, are presented. The resonance region $\sqrt{M^{2}} \le 1$ GeV is 
characterised by a Breit-Wigner parameterisation of the f$_{0}$(500) resonance.
The relative strength of these contributions cannot be specified within
the model, and must hence be determined from the analysis of experimental data.
The model presented allows an extension to central production of strangeonia 
and charmonia states by taking into account the direct-channel contribution 
of the respective trajectories. Moreover, this model can be extended to 
lower beam energies where not only Pomeron-Pomeron, but also Pomeron-Reggeon 
and Reggeon-Reggeon diagrams need to be considered.
The result of the presented work is only the starting point for a comprehensive
study of central exclusive production. To make measurable predictions for 
the LHC, all the diagrams shown in Fig. \ref{fig3} must be calculated. 
The results presented here are necessary and essential input for such 
calculations. Anticipating further studies, we 
remind of the possible reference points that can be used as a guide. 
For the absolute value we use the asymptotic value $\sigma_t\approx 1$ mb, 
compatible with both QCD-inspired and phenomenological estimates 
\cite{Wilk, Cies}. 
The Pomeron-Pomeron total cross section depends also on Pomerons' 
virtualities, $t_1,\ t_2$. We ignored this dependence for two reasons: 
First, this dependence is known at best at their high values, where 
perturbative QCD results, such as that of Eq. (\ref{eq:QCD}), may be valid. 
Second, for simplicity, we fix this dependence, including it as part of the 
normalisation factor. Varying the $t$ dependence and the partition 
between $t_1$ and $t_2$ may be attempted to account for by, 
following Eq. (\ref{eq:QCD}), simply dividing Eq. (\ref{eq2}) by 
$\sqrt{t_1t_2}$, this may be true only for high values of $t_i$, 
beyond diffraction.

\section{Acknowledgements}

We thank Risto Orava and Alessandro Papa for discussions. This work is 
supported by the German Federal Ministry of Education and  Research under 
promotional reference 05P15VHCA1. One of us (L.J.) gratefully acknowledges 
an EMMI visiting Professorship at the University of Heidelberg for 
completion of this work.

\newpage

\section*{Appendix A}

Although the $\rho$-a trajectory does not couple to the $PP$ channel, and 
hence does not contribute to the $PP$ total cross section, 
we list it here to present the procedure for extracting the
coefficients c$_n$ in the expansion of the trajectory.
For convenience, we repeat here the Ansatz for the imaginary part
of the trajectory expressed in Eq. (\ref{eq:imag})

\begin{eqnarray}
\Im m\: \alpha(s) = \sum_{n} c_{n} (s-s_{n})^{1/2} 
\big(\frac{s-s_{n}}{s}\big)^{|\Re e\:\alpha(s_{n})|} \theta(s-s_{n}).
\label{eq:imagl}
\end{eqnarray}

The real part $\Re e\:\alpha(s)$  of the trajectory expressed by 
Eq. (\ref{eq:disp}) can be calculated as

\begin{eqnarray*}
\Re e\:\alpha(s) = \alpha(0) +\frac{s}{\sqrt{\pi}}\sum_{n}c_{n} 
\frac{\Gamma(\lambda_n+3/2)}{\Gamma(\lambda_n+2)\sqrt{s_n}} 
\;_{2}F_{1}(1,1/2;\lambda_n +2;\frac{s}{s_n}) \theta(s_n-s)\;\; +
\end{eqnarray*}
\begin{eqnarray}
\hspace{5.cm}\frac{2}{\sqrt{\pi}}\sum_{n}c_{n}
\frac{\Gamma(\lambda_n+3/2)}{\Gamma(\lambda_n+1)}\sqrt{s_n} 
\;_{2}F_{1}(-\lambda_n,1;3/2;\frac{s_n}{s}) \theta(s-s_n). 
\label{eq:real}
\end{eqnarray}

The derivative $\Re e\:\alpha^{'}$(s) can be derived from Eq. (\ref{eq:real})
\begin{eqnarray*}
\Re e\:\alpha^{'}(s) = \frac{1}{\sqrt{\pi}}\sum_{n}c_{n} 
\frac{\Gamma(\lambda_n+3/2)}{\Gamma(\lambda_n+2)\sqrt{s_n}} 
\;_{2}F_{1}(2,1/2;\lambda_n +2;\frac{s}{s_n}) \theta(s_n-s)\;\; +
\end{eqnarray*}
\vspace{-0.4cm}
\begin{eqnarray}
\hspace{5.cm}\frac{4}{3\sqrt{\pi}}\sum_{n}c_{n}
\frac{\Gamma(\lambda_n+3/2)}{\Gamma(\lambda_n)}\frac{s_n^{3/2}}{s^2} 
\;_{2}F_{1}(1-\lambda_n,2;5/2;\frac{s_n}{s}) \theta(s-s_n). 
\label{eq:realp}
\end{eqnarray}

The equations (\ref{eq:imagl}) - (\ref{eq:realp}) are used to 
calculate the parameters 
c$_{\text n}$ of  $\Re e\:\alpha, \Re e\:\alpha^{'}$ and $\Im m\:\alpha$
by a $\chi^{2}$-fit procedure. A linear fit to the real part provides start 
values for  $\alpha$(0) and $\Re e\:\alpha^{'}$, from which start values for 
$\Im m\;\alpha(M^2_R)$ are calculated.

The fit of the parameters c$_{\text n}$ is done in three steps. 
First, the c$_{\text n}$ are fitted to the expression of the imaginary part 
$\Im m\:\alpha$(s) as given in Eq. (\ref{eq:imagl}).
Second, the coefficients $\alpha$(0) and c$_2$ are extracted by 
using the parameterisation of the real part $\Re e\:\alpha$(s) 
as defined by Eq. (\ref{eq:real}). Third, new values are calculated for 
$\Re e\:\alpha (s_n)$, and the imaginary part $\Im m\:\alpha$(s) is updated  
according to Eq. (\ref{eq:imagl}). These three steps can be 
repeated if necessary until convergence of the values $\alpha$(0) and 
c$_{\text n}$ is reached. 

For fitting the $\rho$-a trajectory, we take the same three thresholds 
s$_{\text i}$ as outlined in Ref. \cite{Fiore1}. The lowest value s$_0$ is taken 
as the 2-pion threshold s$_0$= 4m$_{\pi}^{2}$, with the second value s$_{1}$ 
defined by the a$_2$(1320)-$\pi$ threshold, s$_{1}$= 2.12 GeV$^2$. 
The highest threshold s$_2$ is taken as s$_2$ = 30 GeV$^2$.

\begin{center}
\begin{table}[h]
\begin{tabular}{| c | c c || c | c | c ||}
\hline
& I$^{G}$ & J$^{PC}$ & M (GeV) & M$^{2}$ (GeV$^{2}$) &  $\Gamma$ (GeV) \\ 
\cline{1-6}
$\rho$(770) & 1$^{+}$ &1$^{--}$ &0.769$\pm$ 0.001 &0.591$\pm$ 0.001 &0.149$\pm$0.001 \\
$\rho_{3}(1690)$ & 1$^{+}$ &3$^{--}$ &1.688$\pm$0.002 &2.852$\pm$ 0.007 &0.161$\pm$0.010 \\
$\rho_{5}$(2350) & 1$^{+}$ &5$^{--}$ &2.330$\pm$0.035 &5.429$\pm$0.163 &0.400$\pm$0.100 \\ 
a$_{2}(1320)$ & 1$^{-}$ &2$^{++}$ &1.319$\pm$ 0.001 &1.740$\pm$ 0.003 &0.105$\pm$0.002\\ 
a$_{4}$(2040) & 1$^{-}$ &4$^{++}$ &1.996$\pm$ 0.010 &3.984$\pm$ 0.040 &0.255$\pm$0.026\\ 
a$_{6}$(2450) & 1$^{-}$ &6$^{++}$ &2.450$\pm$ 0.130 &6.003$\pm$ 0.637 &0.400$\pm$0.250\\ 
\hline
\end{tabular}    
\caption{Parameters of $\rho$- and a-resonances.}
\label{table2}
\end{table}
\end{center}

\begin{figure}[h]
\begin{minipage}[t]{.98\textwidth}
\begin{center}
\begin{overpic}[width=.44\textwidth]{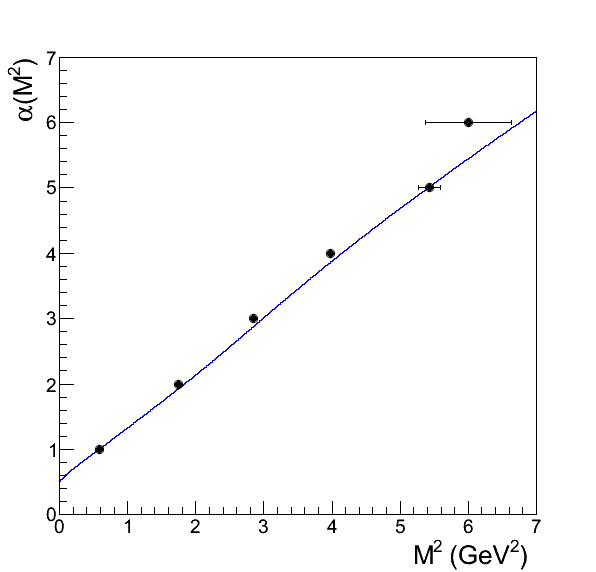}
\end{overpic}
\hspace{.4cm}
\begin{overpic}[width=.44\textwidth]{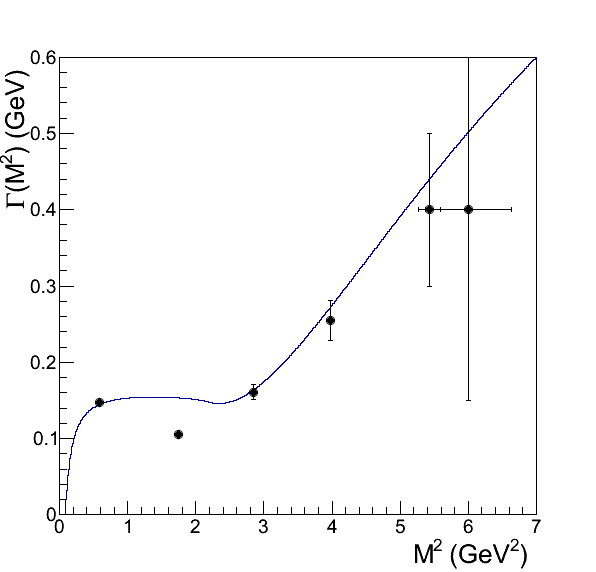}
\end{overpic}
\end{center}
\end{minipage}
\caption{Real part of $\rho$-a trajectory on the left, 
width function $\Gamma$(M$^{2}$) on the right.}
\label{fig:fig2}
\end{figure}

The parameters of the resonances used for the fit of the $\rho$-a trajectory 
are shown in Table \ref{table2}. In Fig. \ref{fig:fig2}, the resulting real 
part and the width function $\Gamma$(M$^{2}$) are shown. 

In the left part of Fig. \ref{fig:fig2}, the black dots represent a plot of
the squared masses M$^2$ of the known resonances $\rho_0(770)$, 
$\rho_3(1670)$, $\rho_5(2350)$, $a_2(1320)$, $a_4(2040)$ and $a_6(2450)$ 
versus their spin. The seemingly linear correlation between these two 
variables, $\alpha$(M$^{2}$) = $\alpha_0$ + $\alpha^{'}$(M$^{2}$), 
is clearly shown by the solid curve as determined 
from \mbox{Eq. (\ref{eq:real}).} For comparison, the star 
symbols superimposed in Fig. \ref{fig:fig2} on the left represent the states 
$\omega_1(782)$, $\omega_3(1670)$, $f_2(1270)$, $f_4(2050)$ and $f_6(2510)$. 
The left part of Fig. \ref{fig:fig2} clearly illustrates the approximate 
degeneracy of the $\rho$-, $\omega$-, $f$- and $a$-trajectory.
The width function $\Gamma$(M$^{2}$) of the $\rho$-a trajectory shown in 
Fig. \ref{fig:fig2} shows good agreement with the corresponding width 
function of Ref. \cite{Fiore1}.


\end{document}